\documentclass[a4paper]{article}

\usepackage[T1]{fontenc}
\usepackage{amsmath}
\usepackage{amssymb}

\usepackage{amsthm}
\newtheorem{theorem}{Theorem}[section]

\newtheorem{claim}[theorem]{Claim}

\theoremstyle{definition}
\newtheorem{definition}{Definition}[section]
\newtheorem{example}{Example}

\theoremstyle{remark}
\newtheorem{remark}{Remark}

\usepackage{xcolor}
\usepackage{todonotes} 

\newcommand{\definenotes}[2]{%
\expandafter\newcommand\csname #1\endcsname[1]{{\color{#2}NOTE(#1): ##1}}
\expandafter\newcommand\csname #1margin\endcsname[1]{\todo[textcolor=#2,backgroundcolor=white]{NOTE(#1): ##1}}
}
\definenotes{avi}{blue}
\definenotes{jack}{red}

\usepackage[inline]{enumitem}
\usepackage{hyperref}
\usepackage{listings}
\usepackage{syntax}
\usepackage{algorithm}
\usepackage{algpseudocode}
\usepackage{tikz}
\usetikzlibrary{arrows,shapes.geometric,positioning,shadows,calc,backgrounds}

\definecolor{codegreen}{rgb}{0,0.6,0}
\definecolor{codegray}{rgb}{0.5,0.5,0.5}
\definecolor{codepurple}{rgb}{0.58,0,0.82}
\definecolor{backcolour}{rgb}{0.95,0.95,0.92}

\lstdefinestyle{mystyle}{
    backgroundcolor=\color{backcolour},
    commentstyle=\color{codegreen},
    keywordstyle=\color{magenta},
    numberstyle=\tiny\color{codegray},
    stringstyle=\color{codepurple},
    basicstyle=\ttfamily\small,
    breakatwhitespace=false,
    breaklines=true,
    captionpos=b,
    keepspaces=true,
    numbers=left,
    numbersep=5pt,
    showspaces=false,
    showstringspaces=false,
    showtabs=false,
    tabsize=2
}
\author{Avi Hayoun\\\small{avi.hayoun@gmail.com}\\\small{Snyk}\and Veselin Raychev\\\small{veselin.raychev@insait.ai}\\\small{INSAIT}\\\small{Sofia University}\\\small{Bulgaria}\and Jack Hair\\\small{jack.hair@snyk.io}\\\small{Snyk}}
\date{\today}
\title{Customizing Static Analysis using Codesearch}

\begin{document}
\maketitle

\begin{abstract}
Static analysis is a growing application of software engineering, leading to a range of essential security tools, bug-finding tools, as well as software verification. Recent years show an increase of universal static analysis tools that validate a range of properties and allow customizing parts of the scanner to validate additional properties or ``static analysis rules''. A commonly used language to describe a range of static analysis applications is Datalog. Unfortunately, the language is still non-trivial to use, leading to analysis that is difficult to implement in a precise but performant way. In this work, we aim to make building custom static analysis tools much easier for developers, while at the same time, providing a familiar framework for application security and static analysis experts. Our approach introduces a language called StarLang, a variant of Datalog which only includes programs with a fast runtime by the means of having low time complexity of its decision procedure.
\end{abstract}

\section{Introduction}\label{sec:intro}

Static analysis is a growing application of software engineering, leading to a range of essential security tools~\cite{SnykCode, CodeQL}, bugfinding tools~\cite{SemGrep, SonarSource} as well as software verification~\cite{Astree, SiDRNS18}. The general idea of static analysis is that some properties of a program are validated without running the code, but instead by reasoning about its possible behaviors. In many cases, these tools are built from the ground up for verifying specific properties such as memory safety~\cite{Astree}, concurrency properties, security properties~\cite{ArztRFBBKTOM14}, type safety~\cite{JensenMT09} or others based on different techniques such as abstract interpretation~\cite{CousotC77}, shape analysis~\cite{RepsSW04}, taint analysis~\cite{ArztRFBBKTOM14} and others. Recent years show an increase of universal static analysis tools such as CodeQL~\cite{CodeQL}, SemGrep~\cite{SemGrep}, Snyk Code~\cite{SnykCode}, and SonarSource~\cite{SonarSource} that validate a range of properties and allow customizing parts of the scanner to validate additional properties or ``static analysis rules''.

This level of customization is usually achieved by unifying parts of the static analysis that are common, such as code parsing, value propagation~\cite{SagivRH95}, dataflow analysis~\cite{RepsHS95}, pointer analysis~\cite{Andersen:1994:ProgramAnalysis, SmaragdakisB15}, typestate analysis~\cite{FinkYDRG08}, and taint analysis~\cite{TrippPFSW09}, and allowing the rules to be written using a domain-specific language or calls to an internal library that provide the base analyses.

A commonly used language to describe a range of static analysis applications is Datalog~\cite{AlpuenteFJV10,JordanSS16, ScholzJSW16}. The language is not Turing complete, yet it was shown to be useful for expressing a range of analyses, such as points-to~\cite{SmaragdakisB15}, security analyses~\cite{GrechS17} and others. Unfortunately, the language is still non-trivial to use, leading to analysis that is difficult to implement in a precise but performant way~\cite{AllenSK15, KastrinisS13, TanLMXS21}. As a result, it is often not suitable for allowing end-users that are non-experts to write rules. This has lead to some of the latest tools providing easier-to-use domain-specific languages to describe their static analysis rules~\cite{SemGrep, SnykCode, CodeQL}.

\paragraph{This work} In this work, we aim to make building custom static analysis tools much easier for developers, while at the same time, providing a familiar framework for application security and static analysis experts. Our approach introduces a language called StarLang, a variant of Datalog which only includes programs with a fast runtime by the means of having low time complexity of its decision procedure. This requirement limits the expressiveness of StarLang, but has the benefit that finding matches in code using StarLang (i.e. matching static analysis alarms) is always fast and allows users to see code matches in real-time as they author a rule. In addition to this, our system can provide autocompletion to help make writing rules even faster and easier.

An interesting observation is that despite StarLang being only as expressive as a subset of Datalog, the language still admits a wide range of useful real-world static analyses. A core reason why this is possible and easy is a \emph{templates} language feature that we introduce. Templates allow hiding Datalog recursion while providing useful high-level abstractions, making the queries easier to read while enforcing the usage of the Datalog subset that is fast and efficient.

\begin{example} We demonstrate the power of StarLang with a simple example in which we would like to discover read-after-close violations using static analysis, i.e., cases of reading from a file after closing it. Consider the following simplified piece of code containing three statements that manipulate a file in a JavaScript-like language:

\begin{lstlisting}
    let f = file();  f.close();  f.read();
\end{lstlisting}

In this case, we are looking for a call to \lstinline|read| and need to follow the dataflow of the receiver object of this call (as a convention, the receiver object is sometimes called argument 0). In this case, one can write the following search query:
\begin{lstlisting}
    CallExpression<"read"> and
    HasArg0<DataFlowAfter<Arg0In<CallExpression<"close">>>>
\end{lstlisting}

Following the logic above, the first line of the query matches the call expression to \lstinline|read|. The second line is a sequence of templates that describe the other checked property --- the receiver object has a call to \lstinline|close| before that. To perform this query, Snyk Code builds a graph that represents the dataflow of the queried code snippet. This graph contains the nodes for the call expressions, as well as edges for the dataflow in the code. Then, each of the StarLang queries or subqueries returns nodes in this graph that match their described properties. For example, \lstinline|CallExpression<"read">| will return every node in the graph that describes a call to \lstinline|read|. The other templates \lstinline|HasArg0|, \lstinline|DataFlowAfter| and \lstinline|Arg0In| traverse edges in the graph, but still always return graph nodes that satisfy the given property, e.g., \lstinline|HasArg0| matches on nodes that have receiver objects with the property given in the template. Some of these templates may also expand into recursive unary predicates, e.g., \lstinline|DataFlowAfter| will match even if there are multiple dataflow steps (i.e., multiple edges) that need to be traversed in the dataflow graph.
\end{example}

\paragraph{Alternative approaches} An alternative approach promoted by SemGrep~\cite{SemGrep}, has been to perform syntactic pattern matching that would allow discovering the pattern above. For example, one can essentially use an expression like \lstinline|$1.close();  ...  $1.read()| to find if there is a variable (or other symbol) that had a sequence of \lstinline|close| and \lstinline|read| operations. Unfortunately, this approach is too syntactic in many cases. As opposed to StarLang, that operates on semantic relations such as dataflow, a syntactic matching approach requires that the specific code pattern matches. As a result, the user needs to modify the rule if they would want to match semantically equivalent, but syntactically different alternative code examples such as the following snippet:

\begin{lstlisting}
    function func(param) { param.close(); }
    let f = file();  func(f);  f.read();
\end{lstlisting}

On the other hand, Snyk Code correctly discovers the read-after-close violation in this snippet using its interprocedural analysis and the same, unmodified StarLang query.
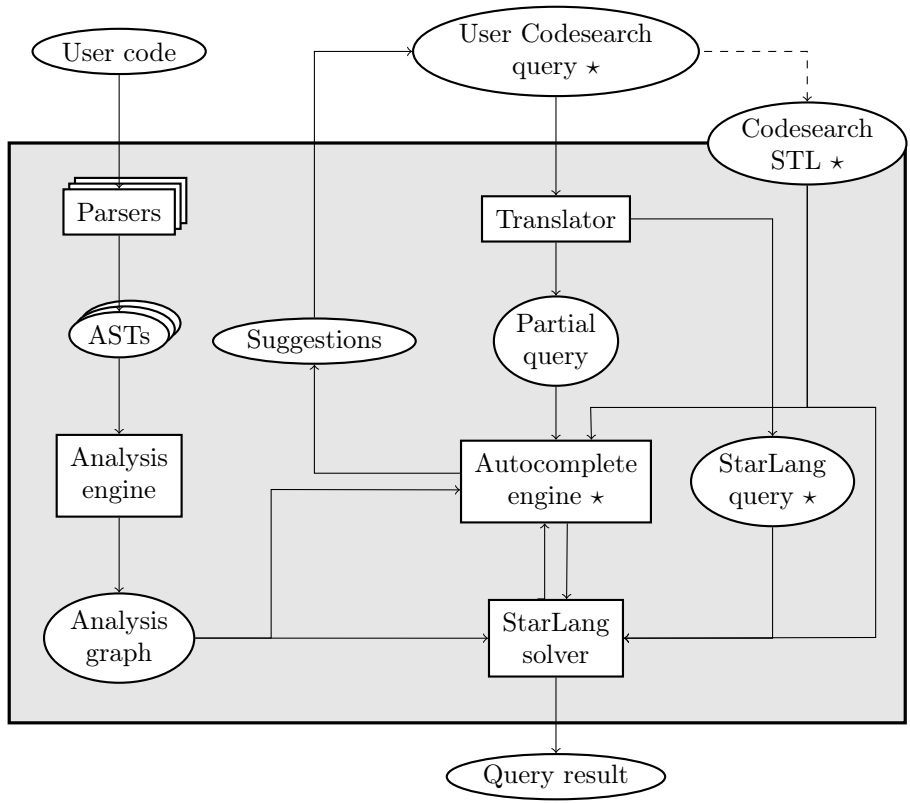
\begin{figure}[!h]
  \begin{tikzpicture}[
    every node/.style={inner sep=1.5pt},
    data/.style={ellipse, draw=black, fill=white, thick, minimum height=4mm, minimum height=6mm, align=center},
    process/.style={rectangle, draw=black, fill=white, thick, minimum width=5mm, minimum height=6mm, align=center, inner sep=5pt},
    shadows/.style={double copy shadow, shadow xshift=2pt, shadow yshift=-2pt},
  ]
    \node[data] (code) {User code};
    \node[data] (query) [right=of code,xshift=1.7cm] {User Codesearch\\query $\star$};
    \node[process, shadows] (parsers) [below=of code,yshift=-0.5cm] {Parsers};
    \node[data, shadows] (asts) [below=of parsers] {ASTs};
    \node[process] (engine) [below=of asts] {Analysis\\engine};
    \node[data] (graph) [below=of engine,yshift=0.1mm] {Analysis\\graph};
    \node[process] (translator) [below=of query,yshift=-0.3cm] {Translator};
    \node[data] (stdlib) [right=of translator,yshift=10mm] {Codesearch\\STL $\star$};
    \node[data] (partial-query) [below=of translator,yshift=0.3cm] {Partial\\query};
    \node[process] (autocomplete) [below=of partial-query,yshift=0.3cm] {Autocomplete\\engine $\star$};
    \node[process] (solver) [below=of autocomplete] {StarLang\\solver};
    \node[data] (starlang-query) [right=of autocomplete,xshift=-5mm] {StarLang\\query $\star$};
    \node[data] (suggestions) [left=of partial-query] {Suggestions};
    \node[data] (result) [below=of solver] {Query result};
  
    \draw[->] (code) -- (parsers);
    \draw[->] (parsers) -- (asts);
    \draw[->] (asts) -- (engine);
    \draw[->] (engine) -- (graph);
    \draw[->] (graph) -- (solver);
    \draw[->] (graph) -| +(2,1.97) -- (autocomplete.185);
    \draw[->] (partial-query) -- (autocomplete);
    \draw[->] (translator) -- (partial-query);
    \draw[->] (query) -- (translator);
    \draw[<-,dashed] (stdlib) |- (query);
    \draw[->] (translator) -| (starlang-query);
    \draw[->] (starlang-query) |- (solver);
    \draw[->] (stdlib) |- +(0.9,-3.5) -- +(0.9,-6.55) -- (solver);
    \draw[->] (stdlib) |- +(0.9,-3.5) -- +(-2.85,-3.5) -- (autocomplete.50);
    \draw[->] (solver) -- (result);
    \draw[->] (solver.115) -| (autocomplete.255);
    \draw[->] (autocomplete.285) -- (solver.75);
    \draw[->] (autocomplete.175) -| (suggestions);
    \draw[->] (suggestions) |- (query);
  
    \begin{scope}[on background layer]
      \draw[very thick,fill=black!10] ($(parsers.north west)+(-0.7,0.6)$) rectangle ($(solver.south east)+(3.7,-0.6)$);
    \end{scope}
  \end{tikzpicture}
  \caption{The Snyk Code Codesearch system. The components in the gray box comprise the system itself, with the inputs and outputs placed above and below respectively. Rectangles represent computations, while ellipses represent data. The ``Codesearch STL'' (STandard Library) node is on the border of the system, acting as an API to the underlying StarLang language.\label{fig:system-design}}
  \end{figure}

\paragraph{Contributions} The contributions of this work are:
    \begin{itemize}
        \item We define StarLang -- a language as expressive as Datalog on unary predicates and show the time complexity of its decision procedures.
        \item We propose a system -- Snyk Code -- that computes dataflow, taint, points-to and other analyses and allows performing StarLang queries in realtime over large repositories. Figure~\ref{fig:system-design} presents a schematic overview of the Snyk Code system.
        \item We demonstrate the expressiveness of StarLang on a range of useful queries that cover taint security properties, memory safety, typestate safety and others.
        \item We present a frontend to StarLang called Codesearch that allows users to interactively built static analysis queries, significantly simplifying the job of security professionals and static analysis authors.
    \end{itemize}

This paper focuses primarily on the parts of Figure~\ref{fig:system-design} marked with a $\star$, and is structured as follows: In Section~\ref{sec:Preliminaries}, we provide an short introduction to Datalog and Relational Algebra. Section~\ref{sec:starlang} formally defines StarLang and the efficiency benefits of its limited expressiveness. In Section~\ref{sec:codesearch} we present Codesearch, a StarLang interface that allows users to define interesting custom static analysis queries. This is demonstrated in Section~\ref{sec:case-studies}, in which we provide examples of such interesting static analysis queries that can be expressed concisely and simply using Codesearch. Finally, Section~\ref{sec:related-work} covers related work.

\section{Preliminaries}\label{sec:Preliminaries}

\subsection{Datalog}
Datalog is a declarative logic programming language that is often used as a query language for deductive databases, and has been employed extensively for program analysis~\cite{AlpuenteFJV10,SmaragdakisKB14,AllenSK15,ScholzJSW16}.

A Datalog program consists of a set of Horn clauses~\cite{Horn51}, called \emph{rules}. Rules are often encoded as implications: \verb|u :- p, q, ..., t.|, where \verb|:-| represents the implication arrow ($\leftarrow$), the commas (,) represent conjunctions ($\wedge$), \verb|.| is the expression delimiter, and \verb|u|, \verb|p|, \verb|q|, etc. are predicates. Logically, a rule is read as: ``if \verb|p, q, ..., t| hold, then \verb|u| holds''. \verb|u| is commonly referred to as the \emph{head} of the rule, while \verb|p, q, ..., t.| is referred to as the rule \emph{body}. An empty rule body is shorthand for a rule body that holds trivially (i.e., is simply \verb|true|). Rules with empty bodies are referred to as \emph{facts}, i.e. they always hold. As usual, predicates are claims of some $n$-ary relation holding for a given set of values, denoted using the standard notation: \texttt{r($X_0$,...,$X_{n-1}$)}, where each $X_i$ is either a concrete value in, or a variable over the universe of relation \verb|r|. Specific predicate instances are sometimes referred to as \emph{atoms}.

Most variants of Datalog support the use of an equality predicate ($=$) over values in the same universe in rule bodies.

A Datalog engine executes the specification for a set of input relations and produces an output relation for a query. The input relations of a Datalog program are referred to as the \emph{Extensional Database} (EDB), and are treated as a set of facts. The rules encoding the logic program are referred to as the \emph{Intensional Database} (IDB).

\begin{example}
    The following Datalog program $P$ computes the ancestor relations relative to an EDB:
    \begin{align*}
        &\texttt{parent(X,Y) :- father(X,Y).} \\
        &\texttt{parent(X,Y) :- mother(X,Y).} \\
        &\texttt{ancestor(X,Y) :- parent(X,Y).} \\
        &\texttt{ancestor(X,Y) :- parent(X,Z), ancestor(Z,Y).}
    \end{align*}

    Consider the following EDB $D$:
    \begin{align*}
        &\texttt{father(john,mary).} \\
        &\texttt{father(joe,kurt).} \\
        &\texttt{mother(mary,joe).} \\
        &\texttt{father(tine,kurt).}
    \end{align*}

    Applying program $P$ to $D$ produces facts such as \verb|ancestor(mary,joe).| and \verb|ancestor(john,joe).|
\end{example}

The basic variant of Datalog does not allow for negating any terms in any rules. While queries of this variant can be solved very efficiently, it has limited use: Clearly, by disabling input negation altogether, one loses
the capability of detecting the absence of database information, which restricts the expressivity of the formalism to queries satisfying some monotonicity property. Extending Datalog to include a limited form of negation, called \emph{stratified negation}~\cite{ChandraH85,GrecoSZ95} (DL$^\neg$) extends the expressiveness, at the cost of higher computational complexity~\cite{DantsinEGV01}. Intuitively, a logic program is a valid program in DL$^\neg$ if it contains no rule that is defined recursively in terms of its own negation (a \emph{negative cycle}). More formally, a program is a valid program in DL$^\neg$ if its IDB rules can be topologically sorted using a straightforward dependency-based ordering: let $r$, and $p$ be two predicate symbols. $p\leq r$ if $p$ appears in a rule body of $r$. If $p$ appears negated in a rule body of $r$, then $p<r$.

\begin{remark}
    Note that this definition admits non-negative recursion: \texttt{$r$ :- $r$.} admits the valid ordering $r\leq r$, but \texttt{$r$ :- $\neg r$.} admits the absurd ordering $r<r$.
\end{remark}

\begin{example}
    The rule \texttt{p(X,Y) :- p(X,Z), $\neg$q(Y, Z).} is a valid DL$^\neg$ rule (assuming \verb|q| is not defined in terms of \verb|p|).
\end{example}


A Datalog program belongs to monadic Datalog (MDL), if all of its rule heads are unary (i.e., have arity 1).

\begin{example}
    The rule \texttt{p(X,Y) :- q(X,Y), r(Y).} is not valid in MDL, since \verb|p(X,Y)| is not unary. However, \texttt{p(X) :- q(X,Y), r(Y).} is a valid MDL rule, given that \verb|q| is an extensional predicate.
\end{example}

\subsubsection{Computational Complexity}
There are several ways to define the complexity of queries or logic programs~\cite{ChandraH82}. They differ in the parameters with respect to which the complexity is measured. The three common complexities in the context of Datalog are:
\begin{itemize}
    \item Data Complexity: the complexity of evaluating a fixed query for variable database inputs.
    \item Program Complexity: the complexity of evaluating, on a fixed database instance, the various queries specifiable.
    \item Combined Complexity: the complexity of evaluating the various queries specifiable for variable database inputs.
\end{itemize}

In the setting of static program analysis, the logic program or query is often pre-determined based on the type of analysis being performed and alarms we would like to detect, and is thus usually considered constant. This means that Data Complexity is the relevant measure in our setting.

\subsection{Relational Algebra}
First introduced by Codd~\cite{Codd70}, relational algebra is a theory that uses algebraic structures for modeling data, and defining queries on it with a well founded semantics~\cite{GelderRS88}. Relational algebra provides a theoretical foundation for relational databases, its main purpose being the definition of operators that transform one or more input relations to an output relation.

\subsubsection{Operators}
Since relations are sets, all of the usual set operations are applicable to them. Additionally, the following operations on relations are defined:

\paragraph{Selection} Denoted with $\sigma_p$, where $p$ is some predicate the domain of which are the elements of the relation to which the selection is being applied. It is used to derive a sub-relation $R'$ from a given relation $R$ such that $R'\subseteq R$ and $\forall r\in R', p(r)$.

\paragraph{Projection} Denoted with $\pi_i$, with $i$ being the index of an entry in the tuples of a given relation $R$. It is used to derive a unary relation $R'$ from $R$ such that $\forall r'\in R', \exists r\in R$ such that the value of $r'$, $x$, is the $i$th entry in $r$: $r=(y_0,\dots,y_i=x,\dots,y_n)$.

\paragraph{Natural Join} Denoted with $\bowtie$. It is used to derive a relation from two given relations, which contains the catenation of tuples that have a matching element at their respective ends: Given relations $R_1$ of $k$-tuples and $R_2$ of $m$-tuples, $R_1\bowtie R_2 = \{(x_0,\dots,x_{k-2},y_0,\dots,y_{m-1}) | (x_0,\dots,x_{k-1})\in R_1 \wedge (y_0,\dots,y_{m-1}) \in R_2 \wedge x_{k-1} = y_0\}$.

The relationship between Datalog and relational algebra is explored in great detail in~\cite{AbiteboulHV95}. In the context of this paper, we make use of the observation that each Datalog rule has an equivalent relational algebra expression. In some cases, it will be simpler and more intuitive to express certain concepts using relational algebraic terms, such as when discussing relation-joining.

\section{StarLang}\label{sec:starlang}
\emph{StarLang} is a strict subset of Monadic Datalog with stratified negation (MDL$^\neg$). As StarLang is a monadic Datalog, every intensional rule defines a unary predicate, i.e., has a single variable in the head of the rule. One of the main restrictions in StarLang is that it does not include the standard explicit equality predicate. Instead, equality is implicit in the reuse of variable names, which is itself restricted.

StarLang rules have a very simple syntactic structure. This is not a semantic limitation. Rather, it is useful for the purposes of presentation, and to make the definition simpler.

\begin{definition}[Valid starlang rules \MakeUppercase{\romannumeral 1}]\label{def:starlang-rules-form}
    A valid StarLang rule $r$ has the form \[\texttt{p(X) :- $\mathcal{A}$, $\mathcal{B}$, $\mathcal{C}$.}\] where $\mathcal{A}$, $\mathcal{B}$ and $\mathcal{C}$ are defined as follows:
    \begin{description}
        \item[$\mathcal{A}$] is a sequence of zero or more predicate citations (either negative or positive) \texttt{$q_1$(X),$\dots$,$q_k$(X)}, where \verb|X| is the variable in the head of $r$.
        \item[$\mathcal{B}$] is an optional pair of conjoined citations. The conjunction, if it is present, has either the form \texttt{$e$(X, Y), $t$(Y)} or the form \texttt{$e$(Y, X), $t$(Y)}, where \verb|X| is the variable in the head of $r$ and \verb|Y|$\neq$\verb|X|. $t$ may be either a positive or a negative citation.
        \item[$\mathcal{C}$] is a sequence of zero or more predicate citations (either negative or positive) \texttt{$d_1$(Z$_1$),$\dots$,$d_j$(Z$_j$)}, where \texttt{Z$_i\neq$X} for $1\leq i\leq j$, \verb|X| being the variable in the head of $r$.
    \end{description}
    An additional limitation is imposed for negative citations which enforces that all negation is stratified: if $q$ is cited negatively in the body of $p$, then $q$ itself must not be defined (directly or transitively) in terms of $p$.
\end{definition}

\begin{example}
    Let \verb|p|, \texttt{e$_1$} and \texttt{e$_2$} be extensional predicates. Then the following is a valid StarLang program:
    \begin{align*}
        &\ \texttt{r(X) :- t$_1$(X), p(X), e$_1$(X, Z), q$_1$(Z), d(U).}\\
        &\texttt{t$_1$(X) :- e$_1$(X, Y), t$_2$(Y).}\\
        &\texttt{t$_2$(Y) :- e$_2$(Y, W), r(W).}\\
        &\texttt{q$_1$(Z) :- e$_2$(V, Z), $\neg$p(V).}
    \end{align*}
\end{example}

\begin{remark}
    Note that Definition~\ref{def:starlang-rules-form} does not allow for n-ary predicates for $n>2$. This is due to the fact that StarLang is a Datalog aimed at static program analysis, and so the input databases would in most cases describe graphs (nodes and edges). Binary predicates naturally describe edges between nodes in a graph, while larger arity predicates are not commonly used to describe graph structures. StarLang could be extended to support hypergraphs (i.e., graphs with hyper-edges), but does not in its current form.
\end{remark}

\begin{remark}
    Definition~\ref{def:starlang-rules-form} does not allow negated binary predicate citation. Since binary predicates in StarLang are assumed to represent directed graph edges, a negated binary predicate would represent a complementary set of edges in the graph. While well-defined and not problematic in general, in the context of static program analysis, this is unlikely to be useful. While StarLang could easily be extended to support edge-predicate negation, it does not in its current form.
\end{remark}

The citations in component $\mathcal{C}$ may seem odd, at first glance. They can, in fact, be very useful, and are necessary for expressing certain queries in StarLang. Due to the limitations on variable-name reuse within a single rule body, without component $\mathcal{C}$, it would be impossible to express constraints on ``disconnected''\footnote{Recall that StarLang was designed to be used as part of a static-analysis engine, and thus the EDB relations are expected to represent forests of program analysis graphs.} parts of the input database in a single query. We refer to the citations in component $\mathcal{C}$ as \emph{disconnected citations}.

Definition~\ref{def:starlang-rules-form} may seem overly restrictive. In fact, it does not encompass all valid StarLang rules. The astute reader might ask about deriving equivalent rules via inlining. And indeed, intuitively, any rule derived by inlining a set of StarLang rules must itself be a valid StarLang rule (as long as proper care is taken around variable naming), due to semantic equivalence. Thus an alternative, more complete definition may be formulated inductively:

\begin{definition}[Valid StarLang rules \MakeUppercase{\romannumeral 2}]\label{def:starlang-rules-II}
    ~\begin{enumerate}
        \item A monadic rule with an empty body, e.g., \texttt{p(X) :- .}, is a valid rule.\label{item:true-valid}
        \item Let $r$ be a valid rule with head $h$ and body $b$, and let \verb|p| be a unary predicate. Then \texttt{$h$ :- $b$, p(X).} is a valid rule.\label{item:add-positive-unary-predicate}
        \item Let $r$ be a valid rule with head $h$ and body $b$, and let \verb|p| be a unary predicate. Then and \texttt{$h$ :- $b$, $\neg$p(X).} is a valid rule, if the rule defining \verb|p| does not depend on the predicate in $h$\footnote{The stratifiability ``check'' here makes this inductive definition not strictly syntactic in its current form. However, one can easily extend this definition to a pair of program and rule $\langle \mathbb{P}, r\rangle$, and the syntactic check would be: make sure that if $p\in\mathbb{P}$ then $r$ is not in the body of $p$ or any of its dependencies.}. \label{item:add-negative-unary-predicate}
        \item Let $r$ be a valid rule with head $h$ and body $b$, and let \verb|e| be an extensional binary predicate. Then \texttt{$h$ :- $b$, e(X, Y).} and \texttt{$h$ :- $b$, e(Y, X).} are valid rules, if \verb|X| is not a fresh variable, but \verb|Y| is (i.e., \verb|X| already appears in $r$ but \verb|Y| does not).\label{item:add-binary-predicate}
    \end{enumerate}
\end{definition}

\begin{remark}
  Items~\ref{item:true-valid}, \ref{item:add-positive-unary-predicate} and \ref{item:add-negative-unary-predicate} of Definition~\ref{def:starlang-rules-II} admit both components $\mathcal{A}$ and $\mathcal{C}$ of Definition~\ref{def:starlang-rules-form}, while items~\ref{item:true-valid} and \ref{item:add-binary-predicate} admit component $\mathcal{B}$: Citation \texttt{$t$(Y)} of component $\mathcal{B}$ may always trivially be defined with an empty rule body.
\end{remark}

\begin{example}
    Let $p_i$ be intensional unary predicates, let $f_j$ be extensional unary predicates, and let $e_k$ be extensional binary predicates.

    The following are valid StarLang rules:
    \begin{itemize}
        \item \texttt{r(X) :- .}
        \item \texttt{r(X) :- f$_1$(X), p$_1$(X), f$_2$(X), p$_2$(X), p$_3$(X), f$_3$(Z).}
        \item \texttt{r(X) :- $\neg$p$_1$(X), f$_1$(X), p$_2$(X).} (assuming \texttt{p$_1$} does not rely on $r$, neither directly nor transitively.)
        \item \texttt{r(X) :- $\neg$p$_1$(X), p$_1$(X).} (trivially equivalent to \verb|false|, but still a valid StarLang rule).
        \item \texttt{r(X) :- e$_1$(X, Y), e$_1$(X, Z), e$_2$(Y, W), r(W), e$_2$(V, Z), p$_2$(V).}
    \end{itemize}

    The following are invalid StarLang rules:
    \begin{itemize}
        \item \texttt{r(X) :- e(Y, X), $\neg$r(Y).} violates item (\ref{item:add-negative-unary-predicate}) --- \verb|r| is negated in the body of \verb|r|.
        \item \texttt{r(X) :- e(Y, Z).} violates item (\ref{item:add-binary-predicate}) --- both \verb|Y| and \verb|Z| are fresh.
        \item \texttt{r(X) :- e$_1$(X, Y), e$_2$(X, Y).} violates item (\ref{item:add-binary-predicate}) --- neither \verb|Y| nor \verb|X| is fresh in \texttt{e$_2$}.
    \end{itemize}
\end{example}

\subsection{Equivalence of Definitions \ref{def:starlang-rules-form} and \ref{def:starlang-rules-II}}
One can use either of the definitions, \ref{def:starlang-rules-form} or \ref{def:starlang-rules-II}, and derive the other. Intuitively, an MDL$^\neg$ rule $r$ is also a valid StarLang rule if it can be transformed into a semantically-equivalent set of rules, each having the ``simple'' $\mathcal{A,B,C}$ structure. What follows is the description of such a procedure. The idea is to construct an undirected graph with labeled edges from a given rule, based on the argument names used in the citations: all unary citations with the same argument will be grouped together in the same node, and binary predicates will define the labeled edges between these nodes (the predicate name being the label). Due to item~\ref{item:add-binary-predicate} of Definition~\ref{def:starlang-rules-II}, the graph structure must be a tree. We then derive a separate rule for each node (structure $\mathcal{A}$ of Definition~\ref{def:starlang-rules-form}) and for each edge (structure $\mathcal{B}$ of Definition~\ref{def:starlang-rules-form}):

\subsubsection{Procedure for transforming a Definition~\ref{def:starlang-rules-II} rule into a Definition~\ref{def:starlang-rules-form} rule.}\label{alg:transform}
Let $r$ be the rule being transformed, let $x$ be the variable in the head of $r$, and let $B$ be the set of predicate citations in the body of $r$. Let $V=\{U(y) | U\subseteq B \wedge \forall u\in B:\text{if }u\text{ is unary with variable }y\text{ then }u\in U\}$. Let $E=\{b(y,z) | b\in B \wedge b\text{ is binary with variables }y,z\}$. Let $G=(V,E)$ be the graph induced by nodes $V$ and edges $E$\footnote{Note that $E$ is not a multi-graph, since item~(\ref{item:add-binary-predicate}) of Definition~\ref{def:starlang-rules-II} prohibits the existence of two binary citations with the same two variables}.

If $V$ contains no node for $x$, then add a node $U(x)=\{\}$ to $V$.

\begin{algorithm}[!hb]
\caption{Translate a rule $r$ matching Definition~\ref{def:starlang-rules-II} into a rule matching Definition~\ref{def:starlang-rules-form} by DFS-traversing the graph induced by $r$}\label{alg:translate}
\begin{algorithmic}
\Procedure{Translate}{Graph $G=\{V, E\}$, Entry point $U(x)$}
\State $stack := V$ where $U(x)$ is at the top of the stack
\State $result := \varnothing$ \Comment{Map of rule heads to rule bodies}
\While {$stack$ is not empty}
\State $top := $ pop top of $stack$
\If {$top$ has not been visited}
\State $y := \text{Var}(top)$
\State $result[t_y] := \langle\rangle$
\If{$s_y$ is a rule in $result$}
\State $result[s_y].\text{add}(t_y(y))$
\ElsIf{$y \neq x$} \Comment{Handle disconnected rules}
\State $result[t_x].\text{add}(t_y(y))$
\EndIf
\For{$b(i,j)$ in edges of $top$}
\If{$b(i,j)$ is incoming (i.e., $i\neq y$) and $i$ is already visited}
\State skip\Comment{Do not add incoming edge rules twice}
\EndIf
\State $z:= (j$ if $i = y$ else $i)$
\State $result[s_z] := \langle b(y,z)\rangle$
\State $result[t_y].add(s_z(y))$
\State push $U(z)$ onto $stack$
\EndFor
\EndIf
\EndWhile
\State\Return $result$
\EndProcedure
\end{algorithmic}
\end{algorithm}

Algorithm~\ref{alg:translate} DFS-traverses $G$ starting at $U(x)$, the node for the variable $x$ in the head of $r$. Nodes of $G$ induce rules $t_u(u)$ for variable $u$ and with fresh predicate name $t_u$. Labeled edges induce rules $s_v(v)$ for variable $v$ and with fresh predicate name $s_v$.

\begin{example}
    Consider the following MDL$^\neg$ rule that matches Definition~\ref{def:starlang-rules-II}:
    \begin{align*}
      \texttt{r(X) :- }&\texttt{e$_1$(X, Y), e$_1$(X, Z), e$_2$(Y, W), r(W), e$_2$(V, Z), $\neg$p(V),}\\
      &\texttt{d(U), e$_1$(U, R).}
    \end{align*}
    It can be trivially transformed into a semantically-equivalent set of rules structured as described above as follows:
    \[ G =
    \begin{cases}
        V =& \{U(\texttt{X}) = \{\}, U(\texttt{W}) = \{\texttt{r(W)}\}, U(\texttt{V}) = \{\neg\texttt{p(V)}\}, U(\texttt{U}) = \{\texttt{d(U)}\}\} \\
        E =& \{\texttt{e$_1$(X, Y), e$_1$(X, Z), e$_1$(U, R), e$_2$(Y, W), e$_2$(V, Z)} \}
    \end{cases}
    \]
    \begin{align*}
        &\texttt{t$_1$(X) :- s$_1$(X), s$_2$(X), t$_5$(U).}\\
        &\texttt{s$_1$(X) :- e$_1$(X, Y), s$_3$(Y).}\\
        &\texttt{s$_2$(X) :- e$_1$(X, Z), s$_4$(Z).}\\
        &\texttt{s$_3$(Y) :- e$_2$(Y, W), t$_2$(W).}\\
        &\texttt{s$_4$(Z) :- e$_2$(V, Z), t$_3$(V).}\\
        &\texttt{t$_2$(W) :- r(W).}\\
        &\texttt{t$_3$(V) :- $\neg$p(V).}\\
        &\texttt{t$_5$(U) :- d(U), s$_5$(U).}\\
        &\texttt{s$_5$(U) :- e$_1$(U, R).}\\
        &\texttt{ r(X) :- t$_1$(X).}
    \end{align*}
\end{example}

\begin{remark}
    As the example demonstrates, Procedure~\ref{alg:transform} generates redundant rules in many cases. This is in order to keep the decision process of the procedure simpler for ease of explanation, not a requirement of StarLang.
\end{remark}

The Procedure~\ref{alg:transform} described above will not produce correct results when starting from an MDL$^\neg$ rule that is not a valid StarLang rule according to Definition~\ref{def:starlang-rules-II}, as the following example demonstrates.

\begin{example} \label{ex:non-equivalent-transformation}
    Consider the MDL$^\neg$ rule that does not match Definition~\ref{def:starlang-rules-II}: \texttt{r(X) :- e$_1$(X, Y), e$_2$(X, Y).}
    It cannot be transformed into a semantically-equivalent set of rules structured as described above; if we apply Procedure~\ref{alg:transform} to this rule, we get the following set of rules:
    \begin{align*}
        &\texttt{t$_1$(X):- s$_1$(X), s$_2$(X).}\\
        &\texttt{s$_1$(X):- e$_1$(X, Y).}\\
        &\texttt{s$_2$(X):- e$_2$(X, Y).}\\
        &\texttt{ r(X) :- t$_1$(X).}
    \end{align*}
    This set of rules is not equivalent to the original MDL$^\neg$ query, since the two instances of \verb|Y| in \texttt{s$_1$} and \texttt{s$_2$} are independent, which may lead different instantiations of \verb|X| in the query result when evaluating the transformed program compared to the original. For example, consider the EDB \{\texttt{e$_1$(1,2).}, \texttt{e$_1$(3,2).}, \texttt{e$_2$(1,2).}, \texttt{e$_2$(3,7).}\}.

    Evaluating the original program on this input results in $\texttt{X}=\{1\}$, while evaluating the transformed program results in $\texttt{X}=\{1,3\}$.
\end{example}

\begin{claim}\label{claim:equivalence}
    Definitions~\ref{def:starlang-rules-form} and \ref{def:starlang-rules-II} are equivalent.
\begin{proof}[Proof sketch]
~\paragraph{(\ref{def:starlang-rules-form}) $\subseteq$ (\ref{def:starlang-rules-II}).} Let $r$ be a rule as described in Definition~\ref{def:starlang-rules-form}. It is easy to check that $r$ adheres to the structural rules laid out in Definition~\ref{def:starlang-rules-II}.

\paragraph{(\ref{def:starlang-rules-form}) $\supseteq$ (\ref{def:starlang-rules-II}).} Let $r$ be a rule as defined in Definition~\ref{def:starlang-rules-II}. Using Procedure~\ref{alg:transform}, $r$ can be transformed into an equivalent set of rules that adheres to the structure described in Definition~\ref{def:starlang-rules-form}.
\end{proof}
\end{claim}

The proof of Claim~\ref{claim:equivalence} relies on the correctness of Procedure~\ref{alg:transform}, i.e., that if the input is input is a valid StarLang rule according to Definition~\ref{def:starlang-rules-II}, then it outputs a semantically-equivalent set of valid StarLang rules according to Definition~\ref{def:starlang-rules-form}. We now prove this holds.

\begin{claim}
Given a valid StarLang rule according to Definition~\ref{def:starlang-rules-II}, Procedure~\ref{alg:transform} outputs a semantically-equivalent set of valid StarLang rules according to Definition~\ref{def:starlang-rules-form}.
\begin{proof}[Proof sketch]
There are three ways in which Procedure~\ref{alg:transform} could violate the claim:
\begin{enumerate*}[label=(\arabic*)]
  \item making dependant variable instances independent (as in Example~\ref{ex:non-equivalent-transformation})\label{item:dependant-independant},
  \item changing the way disconnected rules constrain the query\label{item:disconnected-constraints},
  \item incorrectly handling variable names\label{item:variable-names}, and
  \item generating a disjunctive rule\label{item:disjunctive-rule}.
\end{enumerate*}

Since the procedure does not change any variables, Item~\ref{item:variable-names} is trivially not an issue. Similarly, Item~\ref{item:disjunctive-rule} is not an issue, since Procedure~\ref{alg:transform} cannot generate disjunctions (see Algorithm~\ref{alg:translate}).

For Item~\ref{item:dependant-independant}, observe that due to Item~\ref{item:add-binary-predicate} of Definition~\ref{def:starlang-rules-II}, graph $G$ of the procedure must be a forest of trees; nodes in the graph represent variables, and by Item~\ref{item:add-binary-predicate} of Definition~\ref{def:starlang-rules-II}, exactly one of the variables in a binary predicate (i.e., edge) citation must be fresh. This means that it would be impossible for $G$ to contain two distinct paths between any two nodes.

Finally, for Item~\ref{item:disconnected-constraints}, observe that disconnected rules constraint the entire rule (i.e., are global constraints), and thus evaluating them at the root of the rule sequence is correct. Actually, since Procedure~\ref{alg:transform} does not generate disjunctive rules, adding the disconnected rule constraints at any ``level'' of the output rule sequence would be valid.
\end{proof}
\end{claim}

\subsection{Expressiveness} \label{sec:expressiveness}
As we saw in Example \ref{ex:non-equivalent-transformation}, StarLang is strictly less expressive than MDL$^\neg$:

\begin{claim}\label{claim:mdl-gt-starlang}
    MDL$^\neg$ queries that establish equality between non-head variable operands of $n$-ary predicates ($n>1$) are not expressible in StarLang.
\end{claim}

\begin{example}
The following MDL$^\neg$ query cannot be expressed in StarLang:
$Q\equiv \texttt{p(X) :- q(X,Y), r(X,Y).}$, where \verb|q| and \verb|r| are EDB predicates.

$Q$ is an invalid StarLang query, as it violates restriction \ref{item:add-binary-predicate}.
\end{example}

A corollary of claim~\ref{claim:mdl-gt-starlang} is that queries such as ``Is there a method $m$ of object $O$ that is called both behind a lock and not behind a lock?''\footnote{This query could be useful to find potential concurrency bugs.} cannot be expressed in StarLang. Despite this, StarLang is still quite expressive. For example, since it supports stratified negation, there are many StarLang queries that are not expressible in Semi-positive Datalog --- namely, any query that includes negations of intensional predicates.

\subsubsection{Templates}
StarLang includes a syntactic feature called ``Templates''. These are somewhat similar to Lisp macros --- each template invocation represents a predicate that the compiler auto-generates with the ``holes'' filled by the arguments of the invocation. Unbounded recursion and nested citation are both allowed in templates.

\begin{example}
    Given two predicates \verb|p| and \verb|q|, the following template generates a logic program equivalent to the statement $\texttt{p}\vee(\neg\texttt{p}\wedge\texttt{q})$:
    \[\texttt{TEMPLATE tmpl(p, q) $\equiv$ t(X) :- p(X). t(X) :- $\neg$p(X), q(X).}\]
    where \texttt{t} is a fresh predicate name.
\end{example}

\begin{example}\label{example:recursive-template}
  Given a binary predicate \verb|e| and a predicate \verb|p|, the following template generates a logic program that computes the transitive closure of \verb|e| limited to paths consisting only of nodes for which \verb|p| holds:
  \[\texttt{TEMPLATE tmpl(p) $\equiv$ t(X) :- p(X), e(X, Y), tmpl(p).}\]
  where \texttt{t} is a fresh predicate name.
\end{example}

Templates do not add expressiveness to StarLang; compilation of a templated logic program terminates if and only if the expanded program is finite, in which case it could have been constructed manually. This is the case even for template recursion, as in Example~\ref{example:recursive-template}. This is due to the memoization in the implementation of the expansion procedure: once a hole has been ``filled'' with a given predicate, other instantiations of the same template with the same predicate do not need to be re-expanded.

Even though no expressiveness is added by them, templates significantly improve the usability of the language by allowing users to define abstractions that lead to less code-duplication, thus reducing program size and simplifying program-writing.

\subsection{Efficiency}
StarLang evaluation can be accomplished in time polynomial in the size of the EDB (as is the case for any stratified Datalog with negation~\cite{DantsinEGV01}), for example by using standard bottom-up, semi-na{\"i}ve evaluation~\cite{Bancilhon85}. However, the structure of StarLang rules can be leveraged to achieve a more efficient evaluation process.

Recall the requirement in item~\ref{item:add-binary-predicate} of Definition~\ref{def:starlang-rules-II}, namely that one of the variables in any binary citation must be fresh. This means that constraints expressed using binary predicates are limited to the existence of elements (or ``paths'') in the (cumulative) natural joins of relations, without the use of selection or projection operators.

\begin{example}
    Consider the input EDB $I=\{$\texttt{e$_1$(1,2).}, \texttt{e$_1$(4,3).}, \texttt{e$_1$(2,4).}, \texttt{e$_2$(1,3).}, \texttt{e$_2$(4,2).}, \texttt{e$_2$(2,4).}$\}$.

    It is possible in StarLang to encode the constraint that there must be a ``path'' \texttt{e$_1;$e$_2$} between two elements (i.e., there exists a tuple in the join of \texttt{e$_1$} and \texttt{e$_2$}). For example, the tuple \verb|(1,2,4)| would be an element in such a join in $I$. In relational algebraic terms: $\exists x : x \in (\texttt{e}_1 \bowtie \texttt{e}_2)$.

    However, it is impossible to encode the constraint that there are two elements that are connected by both \texttt{e$_1$} and \texttt{e$_2$} (i.e., that there exists a tuple in the natural join of \texttt{e$_1$} and the inverse relation of \texttt{e$_2$} such that the first and last elements are equal). For example, StarLang can not express a query predicated on the existence of the pair \verb|(2,4)|, which are connected by both \texttt{e$_1$} and \texttt{e$_2$} in $I$. In relational algebraic terms: $\exists x : x \in \sigma_{1 = 3}(\texttt{e}_1 \bowtie \texttt{e}_2^{-1})$.
\end{example}

The constraints on joins of binary relations expressible in StarLang effectively amount to the existence (or lack thereof) of transitive ``paths''. However, answering transitivity queries correctly does not require actually computing the entire join: by the definition of transitivity, path can be computed cumulatively, ``one step at a time''. This results in there being no need to fully compute joins at all when solving StarLang queries; checking transitivity of relations can be thought of as a sequence of selection and projection operations and single, independent joins between unary and binary relations. Thus, while more general Datalog variants require complex runtime join optimizations or manual annotation of join order by users~\cite{ArchHZSS22} in order to achieve remotely-efficient query-solving~\cite{LeisRGMBKN18}, StarLang queries are efficient to compute by design.

\begin{example}
Consider the StarLang query
\[\texttt{r(X) := e$_1$(X, Y), e$_2$(Y, Z), e$_3$(Z, W), p(W).} \] In relational algebraic terms, it could be expressed as $\sigma_\texttt{p}(\texttt{e}_1 \bowtie \texttt{e}_2 \bowtie \texttt{e}_3)$. Alternatively the same query could be expressed as $\pi_1(\texttt{e}_1 \bowtie \pi_1(\texttt{e}_2 \bowtie \pi_1(\texttt{e}_3 \bowtie \texttt{p})))$. Note that in the second form, joins only occur between a binary relation and a unary relation (the result of a projection or the citation \texttt{p}). As a result, the nesting is not necessary --- the projection effectively ``forgets'' the joined relation at each intermediate step, and so the steps can be thought of as a sequence \[R_1 := \pi_1(\texttt{e}_3 \bowtie p);\ R_2 := \pi_1(\texttt{e}_2 \bowtie R_1);\ R_3 := \pi_1(\texttt{e}_1 \bowtie R_2)\] where $R_3$ is the result of the query.
\end{example}

Let $B$ be a binary relation and $U$ a unary relation, and let $\alpha, \beta$ be sorts. If $U : \alpha \times \beta$ and $B : \alpha$, then joining $B$ with $U$ can be done in time $O(\max(|U|, |B|))$, given that $B$ is indexed by its $\alpha$ values (e.g., if $B$ is represented using a hash table). Note that indexing in this case could be done once, as a pre-processing step, since the binary relation being joined is always extensional, i.e., an input relation. Compare that with performing standard computation, with the full joins between two relations $B_1$, $B_2$ instead, which would be $O(|B_1|\cdot|B_2|)$ in the worst case (and could not generally be efficiently computed ahead of time, e.g. in the case of joining temporary relations). Thus, StarLang programs can be evaluated more efficiently than other MDL$^\neg$ programs.

\subsubsection{Data Complexity of StarLang}
Let $P$ be a StarLang program, let $I$ be a finite input database, and let $T$ be the number of elements in the active domain of $I$. Given that StarLang programs are stratifiable, the computation of $P$ over $I$ using a fixed-point approach will take at most $T$ iterations: the process terminates once no new tuple is added to the output relation, and there are at most $T$ tuples which can be added\footnote{This is because a StarLang query must be unary, and so the resulting output relation may only singleton tuples, i.e., at most the entire active domain $T$ of $I$}. In each iteration of the fixed-point computation, at most $|P|$ joins may be evaluated (exactly $|P|$ if the entire program consists of citations of binary EDB relations). Querying a unary relation $U$ can be done in time $|U|$, and there are at most $|P|$ such queries.

Let $m=\max\{|R||R\in I\}$ and the size of the largest EDB relation. Indexing any EDB relation can be done in $O(m)$ time. Let $k$ be the number of EDB relations. Then the data complexity of a StarLang program is bound from above by $O(km+m|P|T)$ --- $km$ being the time to index the input, and $m|P|T$ being the upper bound on the time to compute the solution, given said index. Since $|P|$ is considered to be constant when considering data complexity we get the upper bound $O(m(k+T))$. Thus, the data complexity of StarLang has a linear dependency on three dimensions of the EDB: 
\begin{enumerate*}[label=(\arabic*)]
  \item the size of the largest relation;
  \item the number of relations; and
  \item the number of elements in the active domain.
\end{enumerate*}


\section{Codesearch}\label{sec:codesearch}
Codesearch is a simplified interface for using StarLang in the Snyk Code system provided to non-expert users for the definition of custom semantic analysis rules supported by the Snyk Code analysis engine. The syntax of Codesearch is much more limited than that of StarLang: Users may only define unnamed, stand-alone rules in the form of what is essentially a single disjunction-of-conjunctions query. This query may only contain invocations of pre-defined templates and citations of pre-defined predicates. These predicates and templates are supplied in the form of the StarLang standard library.

As Codesearch is essentially a StarLang ``API'', the abstraction power of templates makes them perfect to act as the primary building blocks of the standard library. The formal grammar of Codesearch can be found in Appendix~\ref{appdx:grammar}, while the description of the definitions in the standard library can be found in Appendix~\ref{appdx:stdlib}.

Despite the additional syntactic limitations, Codesearch is still very expressive, due to the expressiveness of StarLang in combination with the extensiveness of the standard-library. In Section~\ref{sec:case-studies} we present various case studies, demonstrating some interesting queries expressible in Codesearch.

The Codesearch language and the standard library are designed to be as language-agnostic as possible, aiming to make it easy to express abstract semantic concepts, alleviating the need to think about syntactic structures and concrete code patterns.

\begin{example}\label{example:codesearch-query}
Recall the following example presented in Section~\ref{sec:intro}, in which the following Codesearch query matches two semantically-equivalent, but syntactically different code snippets.

\begin{lstlisting}
    CallExpression<"read"> and
    HasArg0<DataFlowAfter<Arg0In<CallExpression<"close">>>>
\end{lstlisting}
Snippet \#1:
\begin{lstlisting}
    let f = file();  f.close();  f.read();
\end{lstlisting}
Snippet \#2:
\begin{lstlisting}
    function func(param) { param.close(); }
    let f = file();  func(f);  f.read();
\end{lstlisting}

This works by Codesearch enabling users to focus on semantic concepts, such as \emph{dataflow} or \emph{call receivers}, moving structural and syntactic concepts, such as \emph{variable names} or \emph{scoping}, to the backseat.
\end{example}

\subsection{Auto-completion}
\begin{figure}[!ht]
    \includegraphics[scale=0.52]{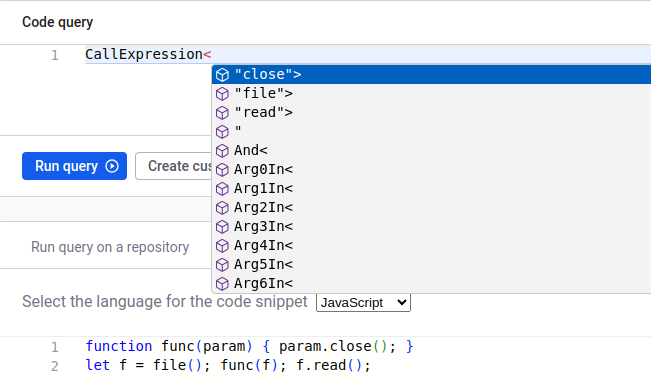}
    \caption{Auto-completion for the argument of the \lstinline|CallExpression| template in the context of the code snippet from Example~\ref{example:codesearch-query}. The most highly-ranked suggestions are all names of functions that are called in the code snippet.\label{fig:autocomplete}}
\end{figure}

To improve the discoverability of standard library definitions, and to assist users in making correct use of said definitions, the Snyk Code Codesearch editor includes an auto-completion mechanism. This mechanism is context-sensitive, and aims to suggest the most relevant literals, predicates and templates, as demonstrated in Figure~\ref{fig:autocomplete}. This context sensitivity is achieved by performing some limited analysis of the code repository being queried, in order to approximate which parts of the code are likely to match for a given template, predicate or literal.

Despite being context sensitive, the auto-completion suggestion system is quite fast. This is done by performing the partial analysis up front and caching the results. This analysis and cache are associated with a single code-base, and not shared among customers. This is both better for customer privacy and data security, as well as better for usability: populating the suggestions based on the analysis of unrelated code repositories would just introduce irrelevant noise to the list of suggestions.

\section{Case Studies}\label{sec:case-studies}
We present a few cases studies, in various programming languages, to demonstrate the expressiveness, conciseness and simplicity of Codesearch.

\subsection{Finding Taint Vulnerabilities}
Taint-flow analysis covers a very large category of security-related vulnerabilities. The following C\# code snippet demonstrates a SQL injection vulnerability, where a string coming from an HTTP request is used to insecurely build a query that is executed on a SQL server.

\begin{lstlisting}[language={[Sharp]C}]
using Microsoft.AspNetCore.Mvc;
using System.Data.SqlClient;

public class DbHandler {
    private const String CONNECTION_STRING = "Server=myServerAddress;Database=myDataBase;User Id=myUsername;Password=myPassword;";

    public static String DoQuery(string query) {
        using (SqlConnection connection = new SqlConnection(CONNECTION_STRING)) {
            SqlCommand cmd = new SqlCommand(query, connection);(*@\label{lst:csharp-sqli}@*)
            return (string)cmd.ExecuteScalar();
        }
    }
}

public class Product { public string? Name { get; set; } }

public class ProductFactory {
    private static String GenQuery(string id) {
        return "SELECT name FROM products WHERE id = " + id;
    }

    public static Product GetProduct(string id) {
        return new Product {
            Name = DbHandler.DoQuery(GenQuery(id))
        };
    }
}

public class VulnerableController : Controller {
    [HttpGet]
    [Route("/product/{id}")]
    public IActionResult ProductInfo(string id) {
        Product product = ProductFactory.GetProduct(id);
        return new JsonResult(product);
    }
}
\end{lstlisting}

The Codesearch standard library includes a dedicated \lstinline|Taint<>| template that allows users to specify a set of sources, sanitizers and sinks to apply to the taint-flow analysis results coming from the analysis engine.

A Codesearch query to match the vulnerability on line~\ref{lst:csharp-sqli} is \lstinline|Taint<PRED:AnySource, PRED:SqliSanitizer, PRED:SqliSink>|. The convenience predicates \lstinline|AnySource|, \lstinline|SqliSanitizer| and \lstinline|SqliSink| are included in the Codesearch standard library as well, along with a multitude of additional taint-flow focused predicates (see Appendix~\ref{appdx:stdlib}).

\subsection{General Dataflow as Taint}
The Codesearch \lstinline|Taint<>| template provides access to more data-flow analysis results than simply taint-flow. As demonstrated in the following example, the \lstinline|Taint<>| template can be used to query non-taint data-flow.
\begin{lstlisting}[language=Java]
package org.example.app;

import java.lang.annotation.*;
import java.lang.String;

@Target(ElementType.FIELD)
@interface Sensitive { }

class User {
    @Sensitive
    private String username;

    User(String username) {
        this.username = username;
    }

    public String getUsername() {
        return username;
    }
}

public class Main {
    public static void main() {
        User user = new User("JohnDoe");
        System.out.println(user.getUsername());(*@\label{lst:java-sensitive-leak}@*)
    }
}
\end{lstlisting}

A Codesearch query that matches on the leak of the \lstinline|Sensitive|-annotated field \lstinline|username| to \lstinline|stdout| on line~\ref{lst:java-sensitive-leak} is \lstinline|Taint<HasAnnotation<"Sensitive">,PRED:None,Arg1In<CallExpression<"java.lang.System.out.println">>>|.

\subsection{Utilizing Typestate Analysis}
Codesearch can be used to detect various typestate analysis-based patterns. The following is an example of use-after-free bugs taken from~\cite{He0S0L22}. This example is a simplification of actual bugs in Internet Explorer (CVE-2010-0249) and the Linux kernel (commit c3aabf0).
\begin{lstlisting}[language=C]
#include <stdlib.h>

typedef struct{ int ref;} A; A* gpA;
typedef struct{ A* _pA;} B; B* gpB;

void dec_ref(A* a) { if (--(a->ref) == 0) free(a); }

void clone(B* b1, B* b2){ b1->_pA = b2->_pA; }(*@\label{lst:c-analysis-gap}\footnotemark@*)

void filename_lookup(A* lpA){ dec_ref(lpA); }

void demo_code(){
    A* pA = (A*)malloc(sizeof(A));
    pA->ref = 1;
    gpA = pA;
    B* pB = (B*)malloc(sizeof(B));
    pB->_pA = pA;
    gpB = (B*)malloc(sizeof(B));
    clone(gpB, pB);
    pB->_pA = NULL;
    free(pB);
    filename_lookup(pA);
}

void main(){
    demo_code();
    dec_ref(gpA);(*@\label{lst:c-uaf-call-1}@*)
    printf(gpB->_pA->ref);(*@\label{lst:c-uaf-call-2}@*)
    free(gpB);
}
\end{lstlisting}
\footnotetext{The definition of \lstinline|clone()| on line~\ref{lst:c-analysis-gap} is slightly different than that in~\cite{He0S0L22}. This is due to a gap in the typestate analysis used performed by the Snyk Code static analysis engine. Given the necessary improvement in the engine, the same Codesearch query would find the bugs with the original definition of \lstinline|clone()|.}

The two bugs are triggered on lines \ref{lst:c-uaf-call-1} and \ref{lst:c-uaf-call-2} (see~\cite{He0S0L22} for a detailed discussion of the bugs). A Codesearch query to match them is \lstinline|DataFlowAfter<Arg1In<CallExpression<free>>>|.

\subsection{Utilizing Points-to Analysis and Negation}
The next example is constructed based on a commit\footnote{\url{https://github.com/apache/lucene-solr/commit/85f5a9c}} from the Apache Lucene project, that fixes a \lstinline|FileInputStream| resource leak. We demonstrate the use of Codesearch to detect points-to analysis patterns.
\begin{lstlisting}[language=Java]
package org.apache.lucene.ant;

import org.w3c.dom.Element;
import org.w3c.tidy.Tidy;

import java.io.FileInputStream;
import java.io.IOException;

public class HtmlDocument {
    private Element rawDoc;

    public HtmlDocumentLeaky(File file) throws IOException {
        Tidy tidy = new Tidy();
        tidy.setQuiet(true);
        tidy.setShowWarnings(false);
        org.w3c.dom.Document root =
            tidy.parseDOM(new FileInputStream(file), null);(*@\label{lst:java-resource-leak}@*)
        rawDoc = root.getDocumentElement();
    }

    public HtmlDocumentFixed(File file) throws IOException {
        Tidy tidy = new Tidy();
        tidy.setQuiet(true);
        tidy.setShowWarnings(false);
        org.w3c.dom.Document root = null;
        InputStream is = new FileInputStream(file);(*@\label{lst:java-resource-no-leak}@*)
        try {
            root =  tidy.parseDOM(is, null);
        } finally {
            is.close();
        }
        rawDoc = root.getDocumentElement();
    }
}
\end{lstlisting}

A Codesearch query that matches on the resource initialization on line \ref{lst:java-resource-leak} but not on \ref{lst:java-resource-no-leak} is \lstinline|CallExpression<"java.io.FileInputStream"> and not ForSameObject<Arg0In<"close">>|. This is also an example of the usefulness of negation in StarLang, which is available in Codesearch.

\subsection{Linting using Syntactic Structures}
The previous examples focus on querying semantic structures based on the results of static analysis. However, as the Codesearch standard library contains templates targeting common syntactic structures (in addition to the many semantically-focused templates), it can be used for simpler tasks as well, such as linting, as demonstrated by the following example.

Default arguments of a function in Python are initialized only once, when the function definition is evaluated. This means that using mutable objects or non-pure expressions as default argument values is probably incorrect. The snippet below demonstrates this with a non-pure function call.

\begin{lstlisting}[language=Python]
from datetime import datetime
import secrets

def gen_filename():(*@\label{lst:python-definition-1}@*)
    return f"{datetime.utcnow().timestamp()}_{secrets.token_hex(16)}.txt"

""""
A possible correct implementation is
def dump_data(data, filename=None):
    if filename is None:
        filename = gen_filename()
        ...
"""
def dump_data(data, filename=gen_filename()):(*@\label{lst:python-definition-2}@*)
    with open(filename, "w") as f:
        f.write(data)

dump_data("foo")(*@\label{lst:python-call-1}@*)
dump_data("bar")(*@\label{lst:python-call-2}@*)
\end{lstlisting}

The function \lstinline[language=Python]{gen_filename} (defined on line~\ref{lst:python-definition-1}) generates a different string each time it is called. However, \lstinline[language=Python]{gen_filename()} is the default value of the argument \lstinline[language=Python]{filename} (line~\ref{lst:python-definition-2}) so the function is only called once. The two calls on lines \ref{lst:python-call-1} and \ref{lst:python-call-2} will write to the same file, with the data from the second call overwriting the data from the first, which is unlikely to be the intended behavior.

A Codesearch query to match on such likely-incorrect default argument definitions is \lstinline|AnyParamIn<DataFlowAfter<CallExpression<*>>>|.

\section{Related Work}\label{sec:related-work}

In this section, we review works closely related to ours.

\paragraph{Restricting Datalog for Efficient Computation} Similar to StarLang, other restricted subsets of Datalog have been found to be useful for efficiently solving a specialized set of problems. Linear Datalog, a restricted form of Datalog that is computable in NL-time, was shown to be equivalent to a class of constraint-satisfaction problems~\cite{Dalmau05}. Symmetric Datalog, a further-restricted form of Linear Datalog that is computable in Logspace, was also utilized in the context of efficiently-computable constraint-satisfaction problems~\cite{EgriLT07}.

\paragraph{Customizing static analysis}
Lint static analysis tools have recently become important components of popular programming languages. While many of these tools start as discovering syntactic stylistic code issues, many expanded into semantic properties and discovering performance, safety and security bugs. For example ESLint~\cite{ESLint}, PyLint~\cite{PyLint}, FindBugs~\cite{FindBugs}, CppCheck~\cite{CppCheck} and StaticCheck~\cite{StaticCheck} are tools for JavaScript, Python, Java, C++ and Go respectively, that currently include dataflow and other analysis as part of their rules. These tools are typically built on top of a common library of analysis, but they still require the rule author to write the code that performs the actual check, only benefiting from reusing the code of the underlying analysis. StarLang is sufficiently expressible to encode most rules of these tools and has large portion of their checks already encoded as Snyk Code quality checks.

Domain specific languages (DSLs) for rules have a number of advantages over manually implemented analyzers, including sharing computation between different rules, better validation of the rules, testing tools and others. Many of these DSL tools, similar to our work, operate on top of graphs to enable easier debugging and visualization. IncA~\cite{szabo2016inca} and CxQL by Checkmarx~\cite{Checkmarx} perform matching based on graph patterns~\cite{graphpatterns}.
PidginQL~\cite{PidginQL} is a query language matching on graphs focusing on enforcing security properties, claiming to perform matching in real-world programs within 15 seconds. Other examples of such tools are SemGrep~\cite{SemGrep} or CodeQL~\cite{CodeQL}. According to the user study in~\cite{piskachev2022fluently}, none of these systems is user-friendly for describing complex security properties such as taint analysis, whereas they present \emph{fluent}TQL~\cite{piskachev2022fluently} that is focused strictly on such taint rules. While all of these tools provide the capability to perform various types of matching on code, none offers time complexity guarantees.

\section{Summary}
We presented StarLang, restricted subset of MDL$^\neg$, which includes only logic programs that can be computed very efficiently, in terms of their data complexity. Based on its efficiency and simplicity characteristics, we constructed an easy-to-use wrapper around StarLang named Codesearch, which allows users to interactively construct static analysis queries, significantly simplifying the job of security professionals and static analysis authors. Finally, we demonstrated the usefulness of Codesearch for its purpose, despite the intentionally-limited expressiveness of both StarLang and Codesearch.

\bibliographystyle{abbrv}
\bibliography{main}

\begin{thebibliography}{10}

\bibitem{Astree}
{Astr\'ee static analyzer}.
\newblock \url{https://www.absint.com/astree/}, 2023.
\newblock [Online; accessed 2023-11-07].

\bibitem{Checkmarx}
{Checkmarx}.
\newblock \url{https://checkmarx.com/}, 2023.
\newblock [Online; accessed 2023-11-03].

\bibitem{CppCheck}
{CppCheck}.
\newblock \url{https://cppcheck.sourceforge.io/}, 2023.
\newblock [Online; accessed 2023-12-15].

\bibitem{ESLint}
{ESLint}.
\newblock \url{https://eslint.org/}, 2023.
\newblock [Online; accessed 2023-12-15].

\bibitem{FindBugs}
{FindBugs}.
\newblock \url{https://findbugs.sourceforge.net/}, 2023.
\newblock [Online; accessed 2023-12-15].

\bibitem{PyLint}
{PyLint}.
\newblock \url{https://pypi.org/project/pylint/}, 2023.
\newblock [Online; accessed 2023-12-15].

\bibitem{SemGrep}
{SemGrep}.
\newblock \url{https://semgrep.dev/}, 2023.
\newblock [Online; accessed 2023-11-03].

\bibitem{SnykCode}
{Snyk Code}.
\newblock \url{https://snyk.io/product/snyk-code/}, 2023.
\newblock [Online; accessed 2023-11-03].

\bibitem{SonarSource}
{SonarSource}.
\newblock \url{https://www.sonarsource.com/}, 2023.
\newblock [Online; accessed 2023-11-03].

\bibitem{StaticCheck}
{StaticCheck}.
\newblock \url{https://staticcheck.dev/}, 2023.
\newblock [Online; accessed 2023-12-15].

\bibitem{AbiteboulHV95}
S.~Abiteboul, R.~Hull, and V.~Vianu.
\newblock {\em Foundations of Databases}.
\newblock Addison-Wesley, 1995.

\bibitem{AllenSK15}
N.~Allen, B.~Scholz, and P.~Krishnan.
\newblock Staged points-to analysis for large code bases.
\newblock In B.~Franke, editor, {\em Compiler Construction - 24th International
  Conference, {CC} 2015, Held as Part of the European Joint Conferences on
  Theory and Practice of Software, {ETAPS} 2015, London, UK, April 11-18, 2015.
  Proceedings}, volume 9031 of {\em Lecture Notes in Computer Science}, pages
  131--150. Springer, 2015.

\bibitem{AlpuenteFJV10}
M.~Alpuente, M.~A. Feli{\'{u}}, C.~Joubert, and A.~Villanueva.
\newblock Datalog-based program analysis with {BES} and {RWL}.
\newblock In O.~de~Moor, G.~Gottlob, T.~Furche, and A.~J. Sellers, editors,
  {\em Datalog Reloaded - First International Workshop, Datalog 2010, Oxford,
  UK, March 16-19, 2010. Revised Selected Papers}, volume 6702 of {\em Lecture
  Notes in Computer Science}, pages 1--20. Springer, 2010.

\bibitem{Andersen:1994:ProgramAnalysis}
L.~Andersen.
\newblock {\em Program Analysis and Specialization for the {C} Programming
  Language}.
\newblock PhD thesis, 1994.
\newblock DIKU Research Report 94/19.

\bibitem{ArchHZSS22}
S.~Arch, X.~Hu, D.~Zhao, P.~Subotic, and B.~Scholz.
\newblock Building a join optimizer for souffl{\'{e}}.
\newblock In A.~Villanueva, editor, {\em Logic-Based Program Synthesis and
  Transformation - 32nd International Symposium, {LOPSTR} 2022, Tbilisi,
  Georgia, September 21-23, 2022, Proceedings}, volume 13474 of {\em Lecture
  Notes in Computer Science}, pages 83--102. Springer, 2022.

\bibitem{ArztRFBBKTOM14}
S.~Arzt, S.~Rasthofer, C.~Fritz, E.~Bodden, A.~Bartel, J.~Klein, Y.~L. Traon,
  D.~Octeau, and P.~D. McDaniel.
\newblock Flowdroid: precise context, flow, field, object-sensitive and
  lifecycle-aware taint analysis for android apps.
\newblock In M.~F.~P. O'Boyle and K.~Pingali, editors, {\em {ACM} {SIGPLAN}
  Conference on Programming Language Design and Implementation, {PLDI} '14,
  Edinburgh, United Kingdom - June 09 - 11, 2014}, pages 259--269. {ACM}, 2014.

\bibitem{CodeQL}
P.~Avgustinov, O.~de~Moor, M.~P. Jones, and M.~Sch{\"{a}}fer.
\newblock {QL:} object-oriented queries on relational data.
\newblock In S.~Krishnamurthi and B.~S. Lerner, editors, {\em 30th European
  Conference on Object-Oriented Programming, {ECOOP} 2016, July 18-22, 2016,
  Rome, Italy}, volume~56 of {\em LIPIcs}, pages 2:1--2:25. Schloss Dagstuhl -
  Leibniz-Zentrum f{\"{u}}r Informatik, 2016.

\bibitem{Bancilhon85}
F.~Bancilhon.
\newblock Naive evaluation of recursively defined relations.
\newblock In M.~L. Brodie and J.~Mylopoulos, editors, {\em On Knowledge Base
  Management Systems: Integrating Artificial Intelligence and Database
  Technologies, Book resulting from the Islamorada Workshop 1985 (Islamorada,
  FL, USA)}, Topics in Information Systems, pages 165--178. Springer, 1985.

\bibitem{ChandraH82}
A.~K. Chandra and D.~Harel.
\newblock Structure and complexity of relational queries.
\newblock {\em J. Comput. Syst. Sci.}, 25(1):99--128, 1982.

\bibitem{ChandraH85}
A.~K. Chandra and D.~Harel.
\newblock Horn clauses queries and generalizations.
\newblock {\em J. Log. Program.}, 2(1):1--15, 1985.

\bibitem{Codd70}
E.~F. Codd.
\newblock A relational model of data for large shared data banks.
\newblock {\em Commun. {ACM}}, 13(6):377--387, 1970.

\bibitem{CousotC77}
P.~Cousot and R.~Cousot.
\newblock Abstract interpretation: {A} unified lattice model for static
  analysis of programs by construction or approximation of fixpoints.
\newblock In R.~M. Graham, M.~A. Harrison, and R.~Sethi, editors, {\em
  Conference Record of the Fourth {ACM} Symposium on Principles of Programming
  Languages, Los Angeles, California, USA, January 1977}, pages 238--252.
  {ACM}, 1977.

\bibitem{Dalmau05}
V.~Dalmau.
\newblock Linear datalog and bounded path duality of relational structures.
\newblock {\em Log. Methods Comput. Sci.}, 1(1), 2005.

\bibitem{DantsinEGV01}
E.~Dantsin, T.~Eiter, G.~Gottlob, and A.~Voronkov.
\newblock Complexity and expressive power of logic programming.
\newblock {\em {ACM} Comput. Surv.}, 33(3):374--425, 2001.

\bibitem{EgriLT07}
L.~Egri, B.~Larose, and P.~Tesson.
\newblock Symmetric datalog and constraint satisfaction problems in logspace.
\newblock In {\em 22nd {IEEE} Symposium on Logic in Computer Science {(LICS}
  2007), 10-12 July 2007, Wroclaw, Poland, Proceedings}, pages 193--202. {IEEE}
  Computer Society, 2007.

\bibitem{FinkYDRG08}
S.~J. Fink, E.~Yahav, N.~Dor, G.~Ramalingam, and E.~Geay.
\newblock Effective typestate verification in the presence of aliasing.
\newblock {\em {ACM} Trans. Softw. Eng. Methodol.}, 17(2):9:1--9:34, 2008.

\bibitem{GelderRS88}
A.~V. Gelder, K.~A. Ross, and J.~S. Schlipf.
\newblock Unfounded sets and well-founded semantics for general logic programs.
\newblock In C.~Edmondson{-}Yurkanan and M.~Yannakakis, editors, {\em
  Proceedings of the Seventh {ACM} {SIGACT-SIGMOD-SIGART} Symposium on
  Principles of Database Systems, March 21-23, 1988, Austin, Texas, {USA}},
  pages 221--230. {ACM}, 1988.

\bibitem{GrechS17}
N.~Grech and Y.~Smaragdakis.
\newblock P/taint: unified points-to and taint analysis.
\newblock {\em Proc. {ACM} Program. Lang.}, 1({OOPSLA}):102:1--102:28, 2017.

\bibitem{GrecoSZ95}
S.~Greco, D.~Sacc{\`{a}}, and C.~Zaniolo.
\newblock {DATALOG} queries with stratified negation and choice: from {P} to
  d\({}^{\mbox{p}}\).
\newblock In G.~Gottlob and M.~Y. Vardi, editors, {\em Database Theory -
  ICDT'95, 5th International Conference, Prague, Czech Republic, January 11-13,
  1995, Proceedings}, volume 893 of {\em Lecture Notes in Computer Science},
  pages 82--96. Springer, 1995.

\bibitem{He0S0L22}
L.~He, H.~Hu, P.~Su, Y.~Cai, and Z.~Liang.
\newblock Freewill: Automatically diagnosing use-after-free bugs via reference
  miscounting detection on binaries.
\newblock In K.~R.~B. Butler and K.~Thomas, editors, {\em 31st {USENIX}
  Security Symposium, {USENIX} Security 2022, Boston, MA, USA, August 10-12,
  2022}, pages 2497--2512. {USENIX} Association, 2022.

\bibitem{Horn51}
A.~Horn.
\newblock On sentences which are true of direct unions of algebras.
\newblock {\em J. Symb. Log.}, 16(1):14--21, 1951.

\bibitem{JensenMT09}
S.~H. Jensen, A.~M{\o}ller, and P.~Thiemann.
\newblock Type analysis for javascript.
\newblock In J.~Palsberg and Z.~Su, editors, {\em Static Analysis, 16th
  International Symposium, {SAS} 2009, Los Angeles, CA, USA, August 9-11, 2009.
  Proceedings}, volume 5673 of {\em Lecture Notes in Computer Science}, pages
  238--255. Springer, 2009.

\bibitem{PidginQL}
A.~Johnson, L.~Waye, S.~Moore, and S.~Chong.
\newblock Exploring and enforcing security guarantees via program dependence
  graphs.
\newblock In D.~Grove and S.~M. Blackburn, editors, {\em Proceedings of the
  36th {ACM} {SIGPLAN} Conference on Programming Language Design and
  Implementation, Portland, OR, USA, June 15-17, 2015}, pages 291--302. {ACM},
  2015.

\bibitem{JordanSS16}
H.~Jordan, B.~Scholz, and P.~Subotic.
\newblock Souffl{\'{e}}: On synthesis of program analyzers.
\newblock In S.~Chaudhuri and A.~Farzan, editors, {\em Computer Aided
  Verification - 28th International Conference, {CAV} 2016, Toronto, ON,
  Canada, July 17-23, 2016, Proceedings, Part {II}}, volume 9780 of {\em
  Lecture Notes in Computer Science}, pages 422--430. Springer, 2016.

\bibitem{KastrinisS13}
G.~Kastrinis and Y.~Smaragdakis.
\newblock Hybrid context-sensitivity for points-to analysis.
\newblock In H.~Boehm and C.~Flanagan, editors, {\em {ACM} {SIGPLAN} Conference
  on Programming Language Design and Implementation, {PLDI} '13, Seattle, WA,
  USA, June 16-19, 2013}, pages 423--434. {ACM}, 2013.

\bibitem{LeisRGMBKN18}
V.~Leis, B.~Radke, A.~Gubichev, A.~Mirchev, P.~A. Boncz, A.~Kemper, and
  T.~Neumann.
\newblock Query optimization through the looking glass, and what we found
  running the join order benchmark.
\newblock {\em {VLDB} J.}, 27(5):643--668, 2018.

\bibitem{piskachev2022fluently}
G.~Piskachev, J.~Sp{\"a}th, I.~Budde, and E.~Bodden.
\newblock Fluently specifying taint-flow queries with fluent tql.
\newblock {\em Empirical Software Engineering}, 27(5):104, 2022.

\bibitem{RepsHS95}
T.~W. Reps, S.~Horwitz, and S.~Sagiv.
\newblock Precise interprocedural dataflow analysis via graph reachability.
\newblock In R.~K. Cytron and P.~Lee, editors, {\em Conference Record of
  POPL'95: 22nd {ACM} {SIGPLAN-SIGACT} Symposium on Principles of Programming
  Languages, San Francisco, California, USA, January 23-25, 1995}, pages
  49--61. {ACM} Press, 1995.

\bibitem{RepsSW04}
T.~W. Reps, S.~Sagiv, and R.~Wilhelm.
\newblock Static program analysis via 3-valued logic.
\newblock In R.~Alur and D.~A. Peled, editors, {\em Computer Aided
  Verification, 16th International Conference, {CAV} 2004, Boston, MA, USA,
  July 13-17, 2004, Proceedings}, volume 3114 of {\em Lecture Notes in Computer
  Science}, pages 15--30. Springer, 2004.

\bibitem{graphpatterns}
G.~Rozenberg, editor.
\newblock {\em Handbook of Graph Grammars and Computing by Graph
  Transformations, Volume 1: Foundations}.
\newblock World Scientific, 1997.

\bibitem{SagivRH95}
S.~Sagiv, T.~W. Reps, and S.~Horwitz.
\newblock Precise interprocedural dataflow analysis with applications to
  constant propagation.
\newblock In P.~D. Mosses, M.~Nielsen, and M.~I. Schwartzbach, editors, {\em
  TAPSOFT'95: Theory and Practice of Software Development, 6th International
  Joint Conference CAAP/FASE, Aarhus, Denmark, May 22-26, 1995, Proceedings},
  volume 915 of {\em Lecture Notes in Computer Science}, pages 651--665.
  Springer, 1995.

\bibitem{ScholzJSW16}
B.~Scholz, H.~Jordan, P.~Subotic, and T.~Westmann.
\newblock On fast large-scale program analysis in datalog.
\newblock In A.~Zaks and M.~V. Hermenegildo, editors, {\em Proceedings of the
  25th International Conference on Compiler Construction, {CC} 2016, Barcelona,
  Spain, March 12-18, 2016}, pages 196--206. {ACM}, 2016.

\bibitem{SiDRNS18}
X.~Si, H.~Dai, M.~Raghothaman, M.~Naik, and L.~Song.
\newblock Learning loop invariants for program verification.
\newblock In S.~Bengio, H.~M. Wallach, H.~Larochelle, K.~Grauman,
  N.~Cesa{-}Bianchi, and R.~Garnett, editors, {\em Advances in Neural
  Information Processing Systems 31: Annual Conference on Neural Information
  Processing Systems 2018, NeurIPS 2018, December 3-8, 2018, Montr{\'{e}}al,
  Canada}, pages 7762--7773, 2018.

\bibitem{SmaragdakisB15}
Y.~Smaragdakis and G.~Balatsouras.
\newblock Pointer analysis.
\newblock {\em Found. Trends Program. Lang.}, 2(1):1--69, 2015.

\bibitem{SmaragdakisKB14}
Y.~Smaragdakis, G.~Kastrinis, and G.~Balatsouras.
\newblock Introspective analysis: context-sensitivity, across the board.
\newblock In M.~F.~P. O'Boyle and K.~Pingali, editors, {\em {ACM} {SIGPLAN}
  Conference on Programming Language Design and Implementation, {PLDI} '14,
  Edinburgh, United Kingdom - June 09 - 11, 2014}, pages 485--495. {ACM}, 2014.

\bibitem{szabo2016inca}
T.~Szab{\'o}, S.~Erdweg, and M.~Voelter.
\newblock Inca: a dsl for the definition of incremental program analyses.
\newblock In {\em Proceedings of the 31st IEEE/ACM International Conference on
  Automated Software Engineering}, pages 320--331, 2016.

\bibitem{TanLMXS21}
T.~Tan, Y.~Li, X.~Ma, C.~Xu, and Y.~Smaragdakis.
\newblock Making pointer analysis more precise by unleashing the power of
  selective context sensitivity.
\newblock {\em Proc. {ACM} Program. Lang.}, 5({OOPSLA}):1--27, 2021.

\bibitem{TrippPFSW09}
O.~Tripp, M.~Pistoia, S.~J. Fink, M.~Sridharan, and O.~Weisman.
\newblock {TAJ:} effective taint analysis of web applications.
\newblock In M.~Hind and A.~Diwan, editors, {\em Proceedings of the 2009 {ACM}
  {SIGPLAN} Conference on Programming Language Design and Implementation,
  {PLDI} 2009, Dublin, Ireland, June 15-21, 2009}, pages 87--97. {ACM}, 2009.

\end{thebibliography}

\newpage
\appendix
\section{Formal Codesearch Grammar Description}\label{appdx:grammar}
\begin{grammar}
    <Query> ::= <Citation> (<LogicalConnective> <Citation>)*

    <Citation> ::= <Negation> (<Predicate> | <Template> | <Literal>)

    <Predicate> ::= `PRED:'<ValidName>

    <Template> ::= <ValidName>`<'<Query>(`,'<Query>)*`>'

    <ValidName> ::= Standard rules for naming variables

    <LogicalConnective> ::= `and' | `or'

    <Negation> ::= `not' | $\varepsilon$

    <Literal> ::= Any UTF-8 character sequence | `~"' RE2 regex `"'
\end{grammar}

In the interest of keeping the grammar description clean, optional grouping of \emph{$\langle$Citation$\rangle$}s using parenthesis has been omitted.

Operator precedence is standard, i.e., \texttt{not} has a higher precedence than \texttt{and}, which has a higher precedence than \texttt{or}.

\section{Codesearch Template and Predicate Library}\label{appdx:stdlib}
\begin{description}
    \item[Predicate \texttt{Any}:] A ``catchall'' rule. Matches on anything.
    \item[Predicate \texttt{AnySink}:] Matches on a range of potential data sinks, including server responses, file systems, database writes, external APIs, logging mechanisms, and other forms of data export or display.
    \item[Predicate \texttt{AnySource}:] Matches on various types of potentially user controlled data sources, both servers (e.g., HTTP parameters/header/body, URLs, cookies, etc.) or indirect ones such as database fields, local files, I/O or environment variables.
    \item[Predicate \texttt{ApexPageReferenceSource}:] Matches on potential XSS sources.
    \item[Predicate \texttt{CleartextCookieStorageSanitizer}:] Matches on cleartext cookie storage sanitizers.
    \item[Predicate \texttt{CleartextCookieStorageSink}:] Matches on cleartext cookie storage sinks.
    \item[Predicate \texttt{CleartextTransmissionSanitizer}:] Matches on cleartext transmission sanitizers.
    \item[Predicate \texttt{CleartextTransmissionSink}:] Matches on cleartext transmission sinks.
    \item[Predicate \texttt{ClientXssSanitizer}:] Matches on client XSS (e.g., DOMXSS) sanitizers.
    \item[Predicate \texttt{ClientXssSink}:] Matches on client XSS (e.g., DOMXSS) sinks.
    \item[Predicate \texttt{CodeInjectionSanitizer}:] Matches on code injection sanitizers.
    \item[Predicate \texttt{CodeInjectionSink}:] Matches on code injection sinks.
    \item[Predicate \texttt{CommandInjectionSanitizer}:] Matches on command injection sanitizers.
    \item[Predicate \texttt{CommandInjectionSink}:] Matches on command injection sinks.
    \item[Predicate \texttt{DeserializationSanitizer}:] Matches on deserialization sanitizers.
    \item[Predicate \texttt{DeserializationSink}:] Matches on deserialization sinks.
    \item[Predicate \texttt{EmailContentInjectionSanitizer}:] Matches on email content injection sanitizers.
    \item[Predicate \texttt{EmailContentInjectionSink}:] Matches on email content injection sinks.
    \item[Predicate \texttt{ErrorMessageOutput}:] Matches on error message outputs (e.g., stacktraces).
    \item[Predicate \texttt{ErrorMessageOutputSanitizer}:] Matches on error message output sanitizers.
    \item[Predicate \texttt{ErrorMessageOutputSink}:] Matches on error message output sinks.
    \item[Predicate \texttt{FileInclusionSanitizer}:] Matches on file inclusion sanitizers.
    \item[Predicate \texttt{FileInclusionSink}:] Matches on file inclusion sinks.
    \item[Predicate \texttt{InformationDisclosureSanitizer}:] Matches on information disclosure sanitizers.
    \item[Predicate \texttt{InformationDisclosureSink}:] Matches on information disclosure sinks.
    \item[Predicate \texttt{JndiInjectionSanitizer}:] Matches on JNDI injection sanitizers.
    \item[Predicate \texttt{JndiInjectionSink}:] Matches on JNDI injection sinks.
    \item[Predicate \texttt{LdapInjectionSanitizer}:] Matches on LDAP injection sanitizers.
    \item[Predicate \texttt{LdapInjectionSink}:] Matches on LDAP injection sinks.
    \item[Predicate \texttt{LogsForgingSanitizer}:] Matches on log-forging sanitizers.
    \item[Predicate \texttt{LogsForgingSink}:] Matches on log-forging sinks.
    \item[Predicate \texttt{MemoryCorruptionSanitizer}:] Matches on prototype memory corruption sanitizers.
    \item[Predicate \texttt{NoSqliSanitizer}:] Matches on NoSQL sanitizers.
    \item[Predicate \texttt{NoSqliSink}:] Matches on NoSQL sinks.
    \item[Predicate \texttt{None}:] An ``anti-catchall'' rule. Matches on nothing.
    \item[Predicate \texttt{OpenRedirectSanitizer}:] Matches on open-redirect sanitizers.
    \item[Predicate \texttt{OpenRedirectSink}:] Matches on open-redirect sinks.
    \item[Predicate \texttt{PointerOperationSink}:] Matches on prototype memory operation sinks.
    \item[Predicate \texttt{PotentialXssSink}:] Matches on potential XSS sinks.
    \item[Predicate \texttt{PrototypePollutionAssignmentSanitizer}:] Matches on prototype pollution assignment sanitizers.
    \item[Predicate \texttt{PrototypePollutionAssignmentSink}:] Matches on prototype pollution assignment sinks.
    \item[Predicate \texttt{PtSanitizer}:] Matches on path-traversal sanitizers.
    \item[Predicate \texttt{PtSink}:] Matches on path-traversal sinks.
    \item[Predicate \texttt{RedosSanitizer}:] Matches on regular-expression denial-of-service sanitizers.
    \item[Predicate \texttt{RedosSink}:] Matches on regular-expression denial-of-service sinks.
    \item[Predicate \texttt{ReflectionSanitizer}:] Matches on reflection sanitizers.
    \item[Predicate \texttt{ReflectionSink}:] Matches on reflection sinks.
    \item[Predicate \texttt{SoqliSanitizer}:] Matches on soqli sanitizers.
    \item[Predicate \texttt{SoqliSink}:] Matches on soqli sinks.
    \item[Predicate \texttt{SosliSanitizer}:] Matches on sosli sanitizers.
    \item[Predicate \texttt{SosliSink}:] Matches on sosli sinks.
    \item[Predicate \texttt{SourceArchive}:] Matches on reading values that are coming from zip, tar or other archives.
    \item[Predicate \texttt{SourceCLI}:] Matches on reading command line arguments.
    \item[Predicate \texttt{SourceClientFramework}:] Matches on reading values that are coming from a client-side framework such as Android, SwiftUI, UIKit, the DOM of an HTML page.
    \item[Predicate \texttt{SourceContainsSensitiveData}:] Matches on reading sensitive data.
    \item[Predicate \texttt{SourceCookie}:] Matches on reading values of cookies in an http server. These values are of security interest, because they can be fully controlled by malicious users.
    \item[Predicate \texttt{SourceDatabase}:] Matches on reading values that are coming from a database.
    \item[Predicate \texttt{SourceEnvironmentVariable}:] Matches on reading environment variables of a process.
    \item[Predicate \texttt{SourceFile}:] Matches on reading values that are coming from files.
    \item[Predicate \texttt{SourceHttpBody}:] Matches on reading http request body in an http server. These values are of security interest, because they may be fully controlled by malicious actors.
    \item[Predicate \texttt{SourceHttpFileUpload}:] Matches on the name and content of file uploaded to an http server. These values are of security interest, because they may be fully controlled by malicious actors.
    \item[Predicate \texttt{SourceHttpHeader}:] Matches on reading values of http headers in a server. These values are of security interest, because they may be fully controlled by malicious actors.
    \item[Predicate \texttt{SourceHttpParam}:] Matches on reading values of http parameters in an http server. These values are of security interest, because they may be fully controlled by malicious actors.
    \item[Predicate \texttt{SourceLocalEnv}:] Matches on reading values from the local environment of the running process. This includes command line arguments, standard input or environment variables.
    \item[Predicate \texttt{SourceNetworkRequest}:] Matches on reading values that are coming from a remote resource through network requests.
    \item[Predicate \texttt{SourceNonServer}:] Matches on reading values that may be controlled by an adversary, but not directly by sending requests to a server. E.g. if an application fetches a value from a URL, an adversary in control of that URL may use it to control its content.
    \item[Predicate \texttt{SourceRequestUrl}:] Matches on reading request URLs in a server. The URLs are of security interest, because they may be fully controlled by malicious actors.
    \item[Predicate \texttt{SourceResourceAccess}:] Matches on reading values that may be controlled by an adversary if they gain access to a resource. The resources this matches are remote URLs, files, database fields or other framework-specific cases such as Android intents.
    \item[Predicate \texttt{SourceRpcApiParam}:] Matches on parameters of RPCs implemented in an RPC server. These values are of security interest, because they may be fully controlled by malicious actors.
    \item[Predicate \texttt{SourceServer}:] Matches on reading values that an attacker can send to a server. Examples are HTTP parameters/header/body, URLs or cookies. Since these may be directly controllable by attacker, these sources are of significant security interest.
    \item[Predicate \texttt{SourceStdin}:] Matches on reading input from the standard input of a process.
    \item[Predicate \texttt{SourceUnrestrictedArchiveFilePath}:] Matches on zipslip sources.
    \item[Predicate \texttt{SourceWebForm}:] Matches on reading values of web forms in a web server. These values are of security interest, because they may be fully controlled by malicious actors.
    \item[Predicate \texttt{SqliSanitizer}:] Matches on SQL injection sanitizers.
    \item[Predicate \texttt{SqliSink}:] Matches on SQL injection sinks.
    \item[Predicate \texttt{SsrfSanitizer}:] Matches on SSRF sanitizers.
    \item[Predicate \texttt{SsrfSink}:] Matches on SSRF sinks.
    \item[Predicate \texttt{SstiSanitizer}:] Matches on SSTI sanitizers.
    \item[Predicate \texttt{SstiSink}:] Matches on SSTI sinks.
    \item[Predicate \texttt{UnsafeSoqliConcatSource}:] Matches on unsafe sosli/soqli concatenations.
    \item[Predicate \texttt{UnsafeSosliConcatSource}:] Matches on unsafe sosli/soqli concatenations.
    \item[Predicate \texttt{XPathInjectionSanitizer}:] Matches on XPath injection sanitizers.
    \item[Predicate \texttt{XPathInjectionSink}:] Matches on XPath injection sinks.
    \item[Predicate \texttt{XamlInjectionSanitizer}:] Matches on XAML injection sanitizers.
    \item[Predicate \texttt{XamlInjectionSink}:] Matches on XAML injection sinks.
    \item[Predicate \texttt{XmlInjectionSanitizer}:] Matches on XML injection sanitizers.
    \item[Predicate \texttt{XmlInjectionSink}:] Matches on XML injection sinks.
    \item[Predicate \texttt{XssSanitizer}:] Matches on XSS sanitizers.
    \item[Predicate \texttt{XssSink}:] Matches on XSS sinks.
    \item[Predicate \texttt{XxeSanitizer}:] Matches on XXE sanitizers.
    \item[Predicate \texttt{XxeSink}:] Matches on XXE sinks.
    \item[Predicate \texttt{ZipSlipSanitizer}:] Matches on zipslip sanitizers.
    \item[Predicate \texttt{ZipSlipSink}:] Matches on zipslip sinks.
  \end{description}
  \begin{description}
    \item[Template \texttt{And}:] A binary conjunction. Matches only if both arguments match. \\
      Has 2 arguments: conjunct, conjunct.
    \item[Template \texttt{AnyParamIn}:] Matches on all parameters of the provided method or function declaration/signature. \\
      Has 1 argument: Function.
    \item[Template \texttt{Arg0In}:] Matches on the 0\textsuperscript{th} index argument (i.e. the receiver object for method calls) for the provided method or function. \\
      Has 1 argument: Function.
    \item[Template \texttt{Arg1In}:] Matches on the 1\textsuperscript{st} index argument for the provided method or function. \\
      Has 1 argument: Function.
    \item[Template \texttt{Arg2In}:] Matches on the 2\textsuperscript{nd} index argument for the provided method or function. \\
      Has 1 argument: Function.
    \item[Template \texttt{Arg3In}:] Matches on the 3\textsuperscript{rd} index argument for the provided method or function. \\
      Has 1 argument: Function.
    \item[Template \texttt{Arg4In}:] Matches on the 4\textsuperscript{th} index argument for the provided method or function. \\
      Has 1 argument: Function.
    \item[Template \texttt{Arg5In}:] Matches on the 5\textsuperscript{th} index argument for the provided method or function. \\
      Has 1 argument: Function.
    \item[Template \texttt{Arg6In}:] Matches on the 6\textsuperscript{th} index argument for the provided method or function. \\
      Has 1 argument: Function.
    \item[Template \texttt{Arg7In}:] Matches on the 7\textsuperscript{th} index argument for the provided method or function. \\
      Has 1 argument: Function.
    \item[Template \texttt{BooleanLiteral}:] Matches on boolean type literals. \\
      Has 1 argument: Value.
    \item[Template \texttt{CallExpression}:] Matches when a given name is called. \\
      Has 1 argument: Function, method or constructor to call.
    \item[Template \texttt{DataFlowAfter}:] Matches on entities that happen after in the dataflow of its parameter. \\
      Has 1 argument: The previous action executed.
    \item[Template \texttt{DataFlowsFrom}:] Matches on places which a taint data can flow from. \\
      Has 1 argument: Source.
    \item[Template \texttt{DataFlowsInto}:] Matches on places which a taint data can flow into. \\
      Has 1 argument: Sink.
    \item[Template \texttt{ExplicitSelfParamIn}:] Matches on the explicit receiver parameter (e.g., \texttt{self} in Python and Rust) for the provided method or function declaration. \\
      Has 1 argument: Function.
    \item[Template \texttt{ForSameObject}:] Matches on entities that happen on the same object as its parameter. \\
      Has 1 argument: The action that happens on the object.
    \item[Template \texttt{HasAnnotation}:] Matches on entities annotated by a given annotation. \\
      Has 1 argument: The annotation with which the entity is annotated.
    \item[Template \texttt{HasAnyArg}:] Matches on entities that take any argument with the provided value. \\
      Has 1 argument: Value.
    \item[Template \texttt{HasArg0}:] Matches on entities that take an argument in the 0\textsuperscript{th} index (i.e. receiver object for method calls) with the provided value. \\
      Has 1 argument: Value.
    \item[Template \texttt{HasArg1}:] Matches on entities that take an argument in the 1\textsuperscript{st} index with the provided value. \\
      Has 1 argument: Value.
    \item[Template \texttt{HasArg2}:] Matches on entities that take an argument in the 2\textsuperscript{nd} index with the provided value. \\
      Has 1 argument: Value.
    \item[Template \texttt{HasArg3}:] Matches on entities that take an argument in the 3\textsuperscript{rd} index with the provided value. \\
      Has 1 argument: Value.
    \item[Template \texttt{HasArg4}:] Matches on entities that take an argument in the 4\textsuperscript{th} index with the provided value. \\
      Has 1 argument: Value.
    \item[Template \texttt{HasArg5}:] Matches on entities that take an argument in the 5\textsuperscript{th} index with the provided value. \\
      Has 1 argument: Value.
    \item[Template \texttt{HasArg6}:] Matches on entities that take an argument in the 6\textsuperscript{th} index with the provided value. \\
      Has 1 argument: Value.
    \item[Template \texttt{HasArg7}:] Matches on entities that take an argument in the 7\textsuperscript{th} index with the provided value. \\
      Has 1 argument: Value.
    \item[Template \texttt{HasNamedArg}:] Matches on entities that take a named argument with the provided value. \\
      Has 2 arguments: The name of the argument., The value the named argument should have.
    \item[Template \texttt{Identifier}:] Matches on an identifier. \\
      Has 1 argument: The entity that should be an identifier.
    \item[Template \texttt{InPath}:] Matches on entities in the source file with the provided path. \\
      Has 1 argument: The path of the file in which to match entities.
    \item[Template \texttt{Literal}:] Matches on string/boolean or number type literals. \\
      Has 1 argument: Value.
    \item[Template \texttt{NamedArgIn}:] Matches on the named argument for the provided method or function. \\
      Has 2 arguments: The name of the argument., The provided method or function.
    \item[Template \texttt{Not}:] A negation. Matches only if the argument does not match. \\
      Has 1 argument: property.
    \item[Template \texttt{NumberLiteral}:] Matches on numeric type literals. \\
      Has 1 argument: Value.
    \item[Template \texttt{Or}:] A binary disjunction. Matches if either (or both) arguments match. \\
      Has 2 arguments: disjunct, disjunct.
    \item[Template \texttt{Param1In}:] Matches on the 1\textsuperscript{st} parameter for the provided method or function declaration. \\
      Has 1 argument: Function.
    \item[Template \texttt{Param2In}:] Matches on the 2\textsuperscript{nd} parameter for the provided method or function declaration. \\
      Has 1 argument: Function.
    \item[Template \texttt{Param3In}:] Matches on the 3\textsuperscript{rd} parameter for the provided method or function declaration. \\
      Has 1 argument: Function.
    \item[Template \texttt{Param4In}:] Matches on the 4\textsuperscript{th} parameter for the provided method or function declaration. \\
      Has 1 argument: Function.
    \item[Template \texttt{Param5In}:] Matches on the 5\textsuperscript{th} parameter for the provided method or function declaration. \\
      Has 1 argument: Function.
    \item[Template \texttt{Param6In}:] Matches on the 6\textsuperscript{th} parameter for the provided method or function declaration. \\
      Has 1 argument: Function.
    \item[Template \texttt{Param7In}:] Matches on the 7\textsuperscript{th} parameter for the provided method or function declaration. \\
      Has 1 argument: Function.
    \item[Template \texttt{ReturnedBy}:] Matches on the returned entity. \\
      Has 1 argument: The entity that returns.
    \item[Template \texttt{Returns}:] Matches on the entity (e.g. a function or a method) that returns the value provided as argument. \\
      Has 1 argument: What is returned.
    \item[Template \texttt{StringLiteral}:] Matches on string type literals. \\
      Has 1 argument: Value.
    \item[Template \texttt{Taint}:] Identify data propagation flows that start at the specified source(s) and reach the designated destination sinks (like vulnerable methods) without going through the specified sanitizer(s). \\
      Has 3 arguments: Source, Sanitizer, Sink.
  \end{description}

  \end{document}